\documentclass[journal=jacsat,manuscript=article]{achemso}

\usepackage{chemformula} 
\usepackage[T1]{fontenc} 
\usepackage{float}
\usepackage{graphicx}
\usepackage{subfigure}
\usepackage{dcolumn}
\usepackage{bm}
\usepackage{hyperref}
\usepackage{natbib}%
\usepackage[compatibility=false]{caption}
\usepackage{xcolor}

\title{Nanometer-scale Cavities for Mid-infrared Radiation via Image Phonon Polariton Resonators}

\author{Michael Klein}
\affiliation{School of Electrical Engineering, Faculty of Engineering, Tel Aviv University, Tel Aviv 6997801, Israel}
\author{Yonatan Gershuni}
\affiliation{School of Electrical Engineering, Faculty of Engineering, Tel Aviv University, Tel Aviv 6997801, Israel}
\author{Alisa Perutski}
\affiliation{School of Electrical Engineering, Faculty of Engineering, Tel Aviv University, Tel Aviv 6997801, Israel}
\author{Jean-Paul Hugonin}
\affiliation{Universite Paris-Saclay, Institut d’Optique Graduate School, CNRS, Laboratoire Charles Fabry, 91127 Palaiseau, France}
\author{Itai Epstein}
\email{itaieps@tauex.tau.ac.il}
\affiliation{School of Electrical Engineering, Faculty of Engineering, Tel Aviv University, Tel Aviv 6997801, Israel}
\alsoaffiliation{Center for Light-Matter Interaction, Tel Aviv University, Tel Aviv 6997801, Israel}
\altaffiliation{QuanTAU, Quantum Science and Technology Center, Tel Aviv University, Tel Aviv 6997801, Israel}

\begin{document}

\begin{abstract}
Surface-polaritons play a pivotal role in strong light-matter interactions at the nanoscale due to their ability to confine light to deep subwavelength dimensions. A promising class of materials exhibiting such polaritonic response are polar dielectrics, which support surface-phonon-polaritons (SPhPs). While SPhPs offer significantly lower losses compared to other polaritons, their potential has been underutilized due to limited ability to reach large continent factors. Here, we demonstrate a system composed of silver nanocubes deposited on a SiC polar dielectric, which experimentally realizes the antisymmetric-image-phonon-polaritons (AIPhPs) mode – a hybridized SPhPs mode that can confine mid-infrared radiation to extremely small mode-volumes, almost a billion times smaller than their free-space volume, with quality-factors an order-of-magnitude greater than those of surface-plasmons or graphene-plasmons, surpassing values of $180$. Our method is general, scalable, and applicable to any polar dielectric, opening the path for controlling and manipulating strong light-matter interactions at the nanoscale in the long-wavelength range.

\end{abstract}

\section{Introduction} 
Surface-polaritons (SPs) are propagating electromagnetic modes that arise from the coupling between electromagnetic radiation and a material excitation in the form of an electric dipole \cite{Maier2007Plasmonics:Applications, Caldwell2015Low-lossPolaritons,Low2016PolaritonsMaterials, Basov2016PolaritonsMaterials}. Fundamentally, they can be described as propagating waves that are confined to the interface between two materials, where one possesses a positive real-part of the permittivity and the other a negative real-part, such as surface-plasmon-polaritons (SPPs) and graphene plasmons (GPs) \cite{Maier2007Plasmonics:Applications, Jablan2009PlasmonicsFrequencies}. When the supporting material is hyperbolic it can give rise to hyperbolic phonon polaritons (HPhPs) with similar properties \cite{Jacob2014HyperbolicPhononpolaritons,Caldwell2014Sub-diffractionalNitride,Dai2014TunableNitride,Fu2024ManipulatingSemiconductor, Eini2022Valley-polarizedFrequencies, Kats2025MonolayerFrequencies}. SPs play an important role in strong light-matter interactions at the nanoscale owing to their ability to confine light to deep subwavelength dimensions, which is accompanied by the enhancement of the optical field \cite{Maier2007Plasmonics:Applications,Ebbesen1998ExtraordinaryArrays,Barnes2003SurfaceOptics,Dionne2005PlanarModel,Dionne2006PlasmonLocalization}. By exploiting these attributes, a wide range of applications have been enabled, such as molecular spectroscopy, narrow-band thermal emitters thin-film sensing, and biosensing \cite{Haas2016AdvancesAnalysis,Berte2018Sub-nanometerPolaritons,Salihoglu2012Plasmon-polaritonsBiosensors,Rodrigo2015Mid-infraredGraphene}. \par

The momentum of SPs can be further increased by placing the polaritonic material near a metallic surface \cite{Moreau2012Controlled-reflectanceNanoantennas, Pisarra2014AcousticGraphene, Alonso-Gonzalez2016AcousticNanoscopy, Chen2017AcousticSpectroscopy, Yuan2020ExtremelyInteractions, Deshpande2018DirectMetasurfaces, Lee2019GrapheneSpectroscopy, Epstein2020Far-fieldVolumes, Lee2020ImageLosses, Menabde2021Real-spaceDeposition, Menabde2022Near-fieldCrystals, Iranzo2018ProbingHeterostructure}. The latter imitates the presence of the polaritonic surface through image charges, and is thus commonly termed image polariton \cite{Lee2020ImageLosses, Menabde2022ImageCrystals, Menabde2021Near-fieldCrystals, Menabde2022Low-Loss-MoO3}. This configuration supports two modes in terms of the electric field distribution: symmetric and anti-symmetric. The symmetric one carries smaller momentum compared to the SP, while the anti-symmetric image polariton (AIP) has been shown to carry extremely large momentum, with confinement factors that are several orders of magnitude smaller then the free-space wavelength  \cite{Hwang2009PlasmonGraphene, Wu2011Large-areaAbsorber, Alonso-Gonzalez2016AcousticNanoscopy, Iranzo2018ProbingHeterostructure, Epstein2020Far-fieldVolumes, Lee2020ImageLosses,Lundeberg2017TuningPlasmonics, Eini2024ElectricallyGraphene,Gershuni2024In-planeLoss, Epstein2020HighlySemiconductors}. \par 

The large momentum carried by SPs, HPhPs and AIPs also introduces a significant momentum mismatch between the polariton and the free-space photon \cite{Ritchie1957PlasmaFilms, Maier2007Plasmonics:Applications, Landa2024ExtendingConversion, Rappoport2020UnderstandingDimensionality}, thus making their excitation and observation challenging. Several approaches have been previously utilized to overcome this issue, such as using patterned structures \cite{ Caldwell2013Low-lossResonators, Zheng2017ExcitationApplications, Iranzo2018ProbingHeterostructure, Lee2020ImageLosses}, or scattering scanning-nearfield-optical microscopy (s-SNOM) \cite{Hillenbrand2002Phonon-enhancedScale, Chen2012OpticalPlasmons, Ni2021Long-LivedMaterials, Ni2018FundamentalPlasmonics, Deshpande2018DirectMetasurfaces, Mancini2022Near-FieldFilms}, and have also been successfully applied to measure surface-plasmon AIPs, graphene-plasmon AIPs  and hyperbolic-phonon-polariton AIPs \cite{Preiner2008EfficientGratings, Moreau2012Controlled-reflectanceNanoantennas, Deshpande2018DirectMetasurfaces, Lee2019GrapheneSpectroscopy, Yuan2020ExtremelyInteractions, Lee2020ImageLosses, Menabde2021Real-spaceDeposition, Menabde2022Near-fieldCrystals, Chen2023Real-spacePolaritons}.\par

Another promising system that exhibits negative permittivity and thus can support SPs is that of polar dielectrics. In these materials, the coupling of charged lattice ions with electromagnetic radiation gives rise to surface-phonon-polaritons (SPhPs) \cite{ Hillenbrand2002Phonon-enhancedScale, Huber2005Near-fieldPropagation, Caldwell2015Low-lossPolaritons, Razdolski2016ResonantNanostructures, Zheng2017ExcitationApplications, Passler2018StrongHeterostructures, Mancini2022Near-FieldFilms}. Since their pioneering realization and measurements by s-SNOM \cite{Hillenbrand2002Phonon-enhancedScale, Huber2005Near-fieldPropagation, Caldwell2013Low-lossResonators, Foteinopoulou2019Phonon-polaritonics:Photonics}, SPhPs have been used in thin-film, biological sensing, and super-resolution imaging \cite{Berte2018Sub-nanometerPolaritons, Neuner2010MidinfraredCarbide, Taubner2006Near-fieldSuperlens}. However, while SPhPs provide inherently low losses \cite{Foteinopoulou2019Phonon-polaritonics:Photonics, Caldwell2015Low-lossPolaritons}, they have not been fully exploited owing to their limited ability to reach large confinement factors compared to other SPs. \par

In this work, we demonstrate extremely confined SPhPs via the experimental realization of their antisymmetric image polariton mode, i.e., the antisymmetric-image-phonon-polaritons (AIPhP). We achieve this by realizing AIPhP resonators in a system composed of nanometric-sized silver cubes deposited on a SiC polar dielectric. We find that the AIPhP resonators can confine midinfrared (MIR) radiation to extremely small mode-volumes, which are almost a billion times smaller than their free-space mode-volume. Moreover, we show that owing to their inherently low losses, AIPhPs exhibit quality factors that are an order of magnitude greater than those of surface-plasmon AIPs and graphene-plasmon AIPs, surpassing values of $180$. Finally, we demonstrate that the spectral response and intensity of the AIPhP resonators can be precisely controlled by adjusting the size and concentration of the nanocubes. \par

\section{Results}
Polar dielectrics, such as SiC for example, exhibit a negative real part permittivity in the spectral range between the transverse optical (TO) and longitudinal optical (LO) phonon frequencies, known as the Reststrahlen band \cite{Caldwell2015Low-lossPolaritons, Foteinopoulou2019Phonon-polaritonics:Photonics}. A signature of this property is the large reflectivity observed in the Reststrahlen band's spectral range, stemming from the material's metal-like behavior. Although SPhPs carry smaller momentum compared to GPs and HPhPs, they can provide reduced propagation losses, thus it is worthwhile to realize their AIP equivalent. \par

\begin{figure}
    \centering
    \includegraphics[width=0.8\linewidth]{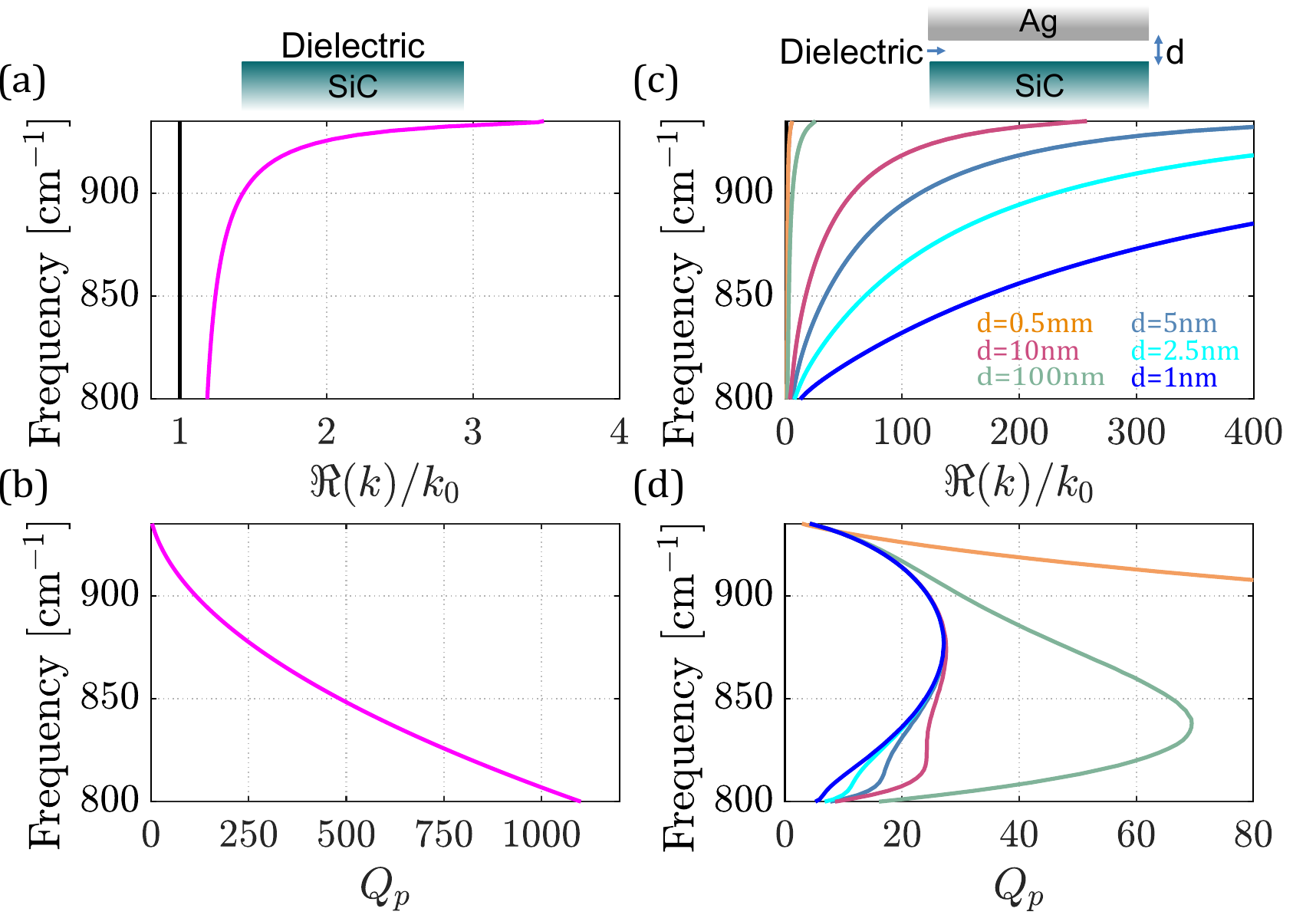}
        \caption{Confinement factor $k/k0$ and loss figure of merit $Q_p$ of SPhPs and AIPhPs.
        (a) Calculated $k/k0$ compared to the lightline (black curve) and (b) $Q_p$ of SPhPs at the interface between semi-infinite SiC and a dielectric within the Reststrahlen band spectral range. (c) Calculated $k/k0$ and (d) $Q_p$ of AIPhPs supported in a SiC/dielectric/Ag structure with a dielectric spacer of thickness $d$. Both geometries are shown at the top part of the figure.
        }
    \label{fig:1}
\end{figure}

To evaluate the optical performance of SPhPs in comparison to AIPhPs, it is convenient to define the confinement factor, $k/k0$, and loss figure of merit, $Q_p = \mathrm{Re}(k_p)/\mathrm{Im}(k_p)$ \cite{Ni2018FundamentalPlasmonics,Woessner2014HighlyHeterostructures}. Figure \ref{fig:1} shows the simulated values (see SI) for SPhPs supported at the interface between semi-infinite SiC and a dielectric ($n=1.4$) (fig. \ref{fig:1}(a),(b)), and those for AIPhPs supported in a SiC/dielectric/Ag structure with a dielectric spacer of thickness $d$ (fig. \ref{fig:1}(c),(d)). It can be seen in Fig. \ref{fig:1}(a),(b) that SPhPs indeed exhibit low confinement factors nearing the lightline, $k_0=\omega/c$ (black curve), accompanied by large values of $Q_p$ owing to their low losses. In contrast, fig. \ref{fig:1}(c),(d) show that AIPhPs realized by bringing the SiC to a distance $d$ from an Ag surface yields extremely large confinement factors, still with relatively low losses. As $d$ decreases, the momentum increases by up to two orders of magnitude, and the dispersion becomes approximately linear, a characteristic behavior of AIP modes \cite{Alonso-Gonzalez2016AcousticNanoscopy,Lundeberg2017TuningPlasmonics, Iranzo2018ProbingHeterostructure, Epstein2020Far-fieldVolumes}. Despite the increased losses at smaller $d$ separations, at optimal conditions ($875 [cm^{-1}]$) the confinement factor exceeds 350 with $Q_p$ values reaching $~27$ for the smallest spacer thickness.


\begin{figure}
    \centering
    \includegraphics[width=0.8\linewidth]{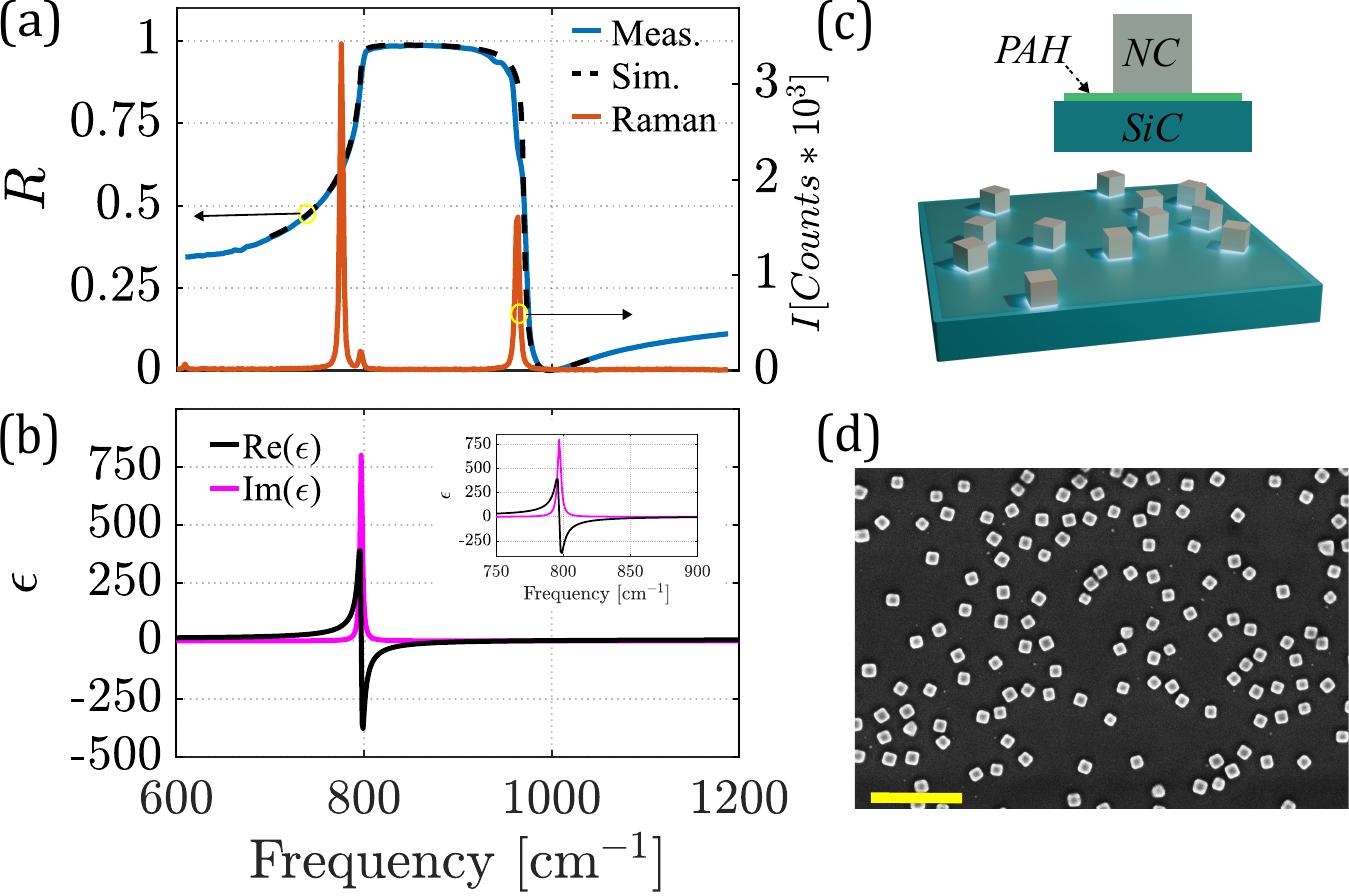}
        \caption{Characterization and analysis of the 4H-SiC substrate. (a) FTIR reflection spectrum (blue curve) and Raman spectrum (red curve) of the 4H-SiC material, showing the Reststrahlen band spectral range and the LO and TO phonons. The spectrum was calculated and fitted (black dashed curve) using Eq.\ref{SiC Dielectric Constant}. (b) Real (black curve) and imaginary (magenta curve) parts of the permittivity of 4H-SiC, showing the correlation between the negative real part of the permittivity to the Reststrahlen band in (a), inset: SiC permittivity around its Reststrahlen band. (c) Sample configuration showing the 4H-SiC substrate, coated with a PAH layer, and covered with nanometer-sized silver cubes. The inset shows the cross-section of a single SiC-NC structure with a thin PAH layer between them. (d) SEM image of $100 nm$ silver NCs randomly dispersed on top of 4H-SiC (scale bar, $600 nm$).}
    \label{fig:2}

\end{figure}

To realize such extremely confined AIPhPs, we start first by characterizing the 4H-SiC crystal at hand. Figure \ref{fig:2}a presents the optical characterization of the SiC crystal in the Reststrahlen band's spectral range via reflection (blue curve) and Raman (red curve) measurements. The Raman curve clearly shows the locations of the TO and LO phonons, and the expected large reflectivity in this range can be observed. The dashed black curve in Fig. \ref{fig:2}a shows the reflection spectrum obtained from the transfer-matrix-method simulation of the structure (see SI), with the permittivity of the SiC shown in Fig. \ref{fig:2}b modeled by the Lorentz model for polar materials \cite{Foteinopoulou2019Phonon-polaritonics:Photonics}:

\begin{align}\label{SiC Dielectric Constant} \epsilon(\omega)&= \epsilon_{\infty}\left(\frac{\omega_{LO}^2-\omega^2-i\Gamma_{LO}\omega}{\omega_{TO}^2-\omega^2-i\Gamma_{TO}\omega}\right) \end{align}

where $\epsilon_{\infty}$ is the high-frequency permittivity; $\omega_{LO}$ and $\omega_{TO},$ are the LO and TO frequencies, respectively, and $\Gamma_{LO}$ and $\Gamma_{TO}$ are the phonon's damping rates. \par

From the simulation fit of Eq.\ref{SiC Dielectric Constant} to the measured reflection spectrum we can extract the $\omega_{LO}=971 cm^{-1}$, $\omega_{TO}= 796 cm^{-1}$, $\Gamma_{LO} = 3 cm^{-1}$ and $\Gamma_{TO}= 2.9 cm^{-1}$ values, where $\epsilon_{\infty}=6.56$ \cite{Harima1995Raman6HSiC} is presented in Fig. \ref{fig:1}b (dashed black curve) and shows good agreement. For completeness, the obtained complex permittivity values are presented in Fig. \ref{fig:2}b. \par

Since it is not practical to bring a metallic surface close to a SiC crystal, for the AIPhPs realization, we cover the bare SiC substrate with a thin layer of polyallylamine hydrochloride (PAH), followed by the deposition of $100 nm$ silver NCs (see methods). Figure \ref{fig:2}c shows an illustration of the device, where NCs are randomly scattered on top of a SiC crystal and a thin layer of PAH between them. The top face of the device can be seen in the scanning electron microscopy image (Fig. \ref{fig:2}d), illustrating the random position and orientation of the silver NCs. \par 

\begin{figure}
 \centering
           \includegraphics[width=1\linewidth]{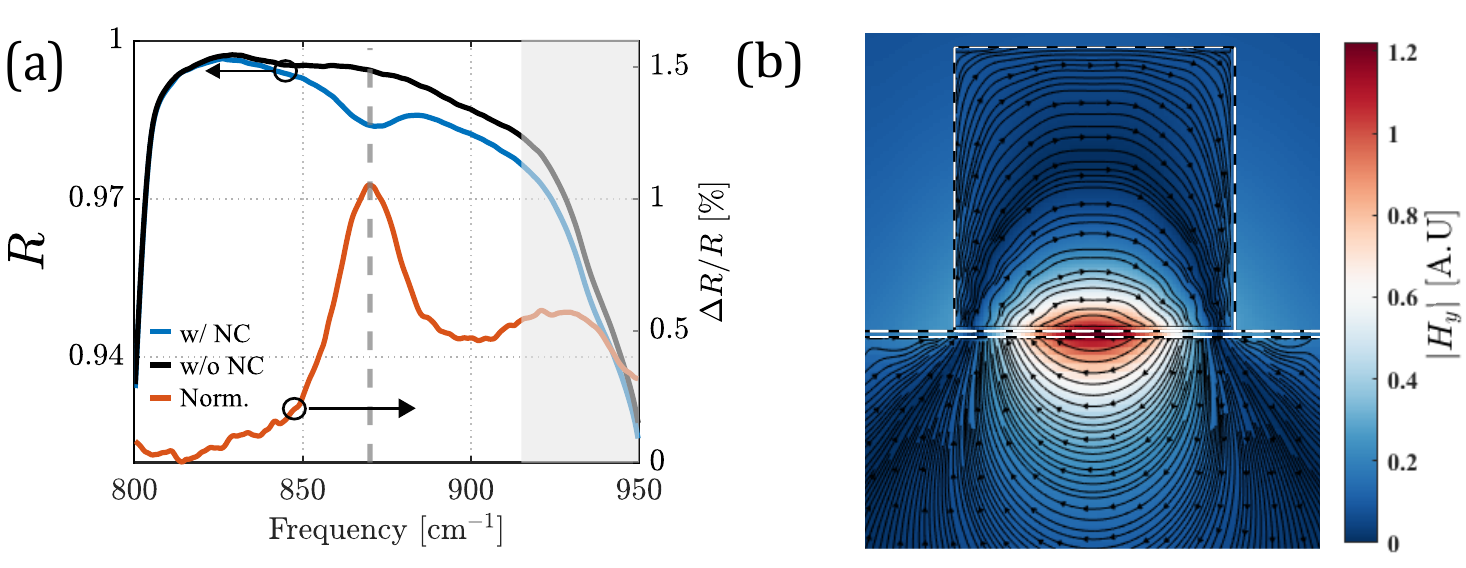}

             \caption{Characterization of the spectra obtained from the device comprising NCs deposited on SiC. (a) FTIR reflection measurements of the SiC substrate with (blue curve) and without (black curve) $100 nm$ silver NCs. The reflection contrast measurement (orange curve) illustrates the isolated resonance response of the structure. The grey dashed vertical line highlights the peak location, and the gray area covers the broader peak resulting from the Reststrahlen band's tail. (b) 3D simulations of the magnetic field distribution over a single resonator, $|Hy|$, superimposed with the electric field lines, showing the generation of a magnetic dipole resonance at the SiC-NC interface at resonance. In these simulations, vacuum was used as the environment and a PAH spacer thickness of $1 nm$.}
             \label{fig:3}
\end{figure}

Next, we analyze the optical response of our device within the spectral range of the Reststrahlen band. The reflection measurements with and without the NCs are shown for comparison in Fig. \ref{fig:3}a. A single dip can be seen at a frequency of $870 cm^{-1}$, marked by the vertical dashed gray line, when the NCs are present (blue curve), which is not observed in the measurement of the bare SiC (black line). In order to clearly separate the resonance from the large reflective background of the Reststrahlen band, we plot in Fig. \ref{fig:3}a the reflectance contrast (orange curve), defined as $\Delta R/R$, where $\Delta R=R-R_{w/NC}$, and $R_{w/NC}$ and $R$ are the reflections measured with and without the NCs, respectively. A distinct resonant peak in the reflection contrast can be observed in Fig. \ref{fig:3}a, at $870 cm^{-1}$ frequency. We note that an additional peak can also be observed at $930 cm^-1$ frequency, indicated by the gray area in Fig. \ref{fig:3}a. This broader peak results from the rapid decrease and small changes in the reflection curves close to the edge of the Reststrahlen band, and is thus not associated with the AIPhPs resonance. \par 

To understand the nature of the observed resonance, we perform three-dimensional simulations of a single SiC-NC resonator excited by far-field radiation (see methods) \cite{Lalanne1996HighlyPolarization, Lalanne2005PerfectlyFormalization}. The magnetic field distribution obtained at the resonance frequency ($870 cm^{-1}$) over the entire structure is presented in Fig. \ref{fig:2}b, superimposed with the electric field lines. It can be seen that the electric field lines form a loop encircling an area with a strong magnetic field at the SiC-NC interface. This behavior is a signature of the AIP, as has been previously observed in surface-plasmon AIPs and graphene-plasmon AIPs \cite{Bozhevolnyi2007GeneralResonators, Moreau2012Controlled-reflectanceNanoantennas, Epstein2020Far-fieldVolumes}. From this, we can conclude that the nature of the observed resonance is the excitation of AIPhPs, where each NC together with the SiC acts as a single AIPhP resonator. The principle of operation is that of a patch antenna, in which a rectangular metallic patch above a metallic surface creates similar magnetic dipole resonances \cite{Moreau2012Controlled-reflectanceNanoantennas, Epstein2020Far-fieldVolumes}, where in our case, the negative permittivity of the SiC acts as a metal-like surface within the Reststrahlen band's spectral range. \par

\begin{figure}
 \centering
           \includegraphics[width=1\linewidth]{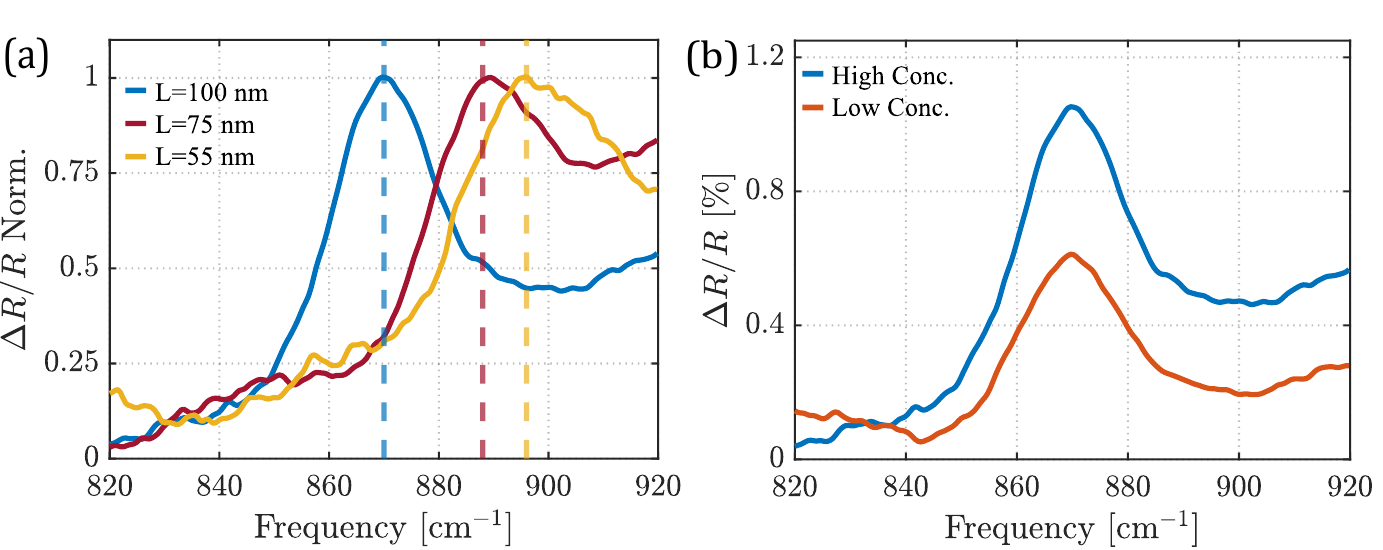}
           
     \caption{Analysis of the change in the spectral response of the APhPs resonators with  - (a) the change in the NCs size ($L= 55nm$, $75nm$ and $100nm$), and (b) the change in the NCs concentration, for same NC size, $L=100nm$. The  APhP resonance peak moves to lower frequencies as the NC size decreases, while the amplitude of the response is only affected by the different concentrations. Each resonant peak is marked with a corresponding dashed vertical line. $L$ is the NC dimensions,
     $\Delta R/R$ {Norm.}=$\Delta R/(R*R_{max})$ }
     \label{fig:4} 
\end{figure}

Next, we examine the spectral response of the AIPhP resonators of different NC sizes and concentrations. The measured normalized reflection of three different samples with NC sizes of $100nm$ (blue curve), $75 nm$ (orange curve), and $55 nm$ (yellow curve) are presented in Fig. \ref{fig:4}a. It is notable that the AIPhP resonance shifts to higher frequencies with decreasing NC size, as expected from the dispersive nature of the AIPhP resonance \cite{Epstein2020Far-fieldVolumes}. To further characterize the optical response, we conducted measurements of the reflection from two additional samples with $100 nm$ cube size but differing NC concentrations (Fig. \ref{fig:4}b), one with approximately twice the NC density of the other (see SI). It can be seen from the measurements that the resonance frequency remains constant in both samples, as it is the NC size that sets the resonance frequency, while the amplitude of the response changes only with concentration, with greater concentration corresponding to a larger amplitude. These results agree well with previous observations of graphene-plasmon AIPs and surface-plasmon AIPs \cite{Epstein2020Far-fieldVolumes, Moreau2012Controlled-reflectanceNanoantennas}. We note that the reflection measurement captures the collective response of many resonators within the beam's spot, resulting in the measurement of the averaged response, stemming from the nanocubes' shape and size variations, together with their possible deformation and  aggregation after deposition on the SiC surface.\par

\begin{figure}

           \centering
          \includegraphics[width=1\linewidth]{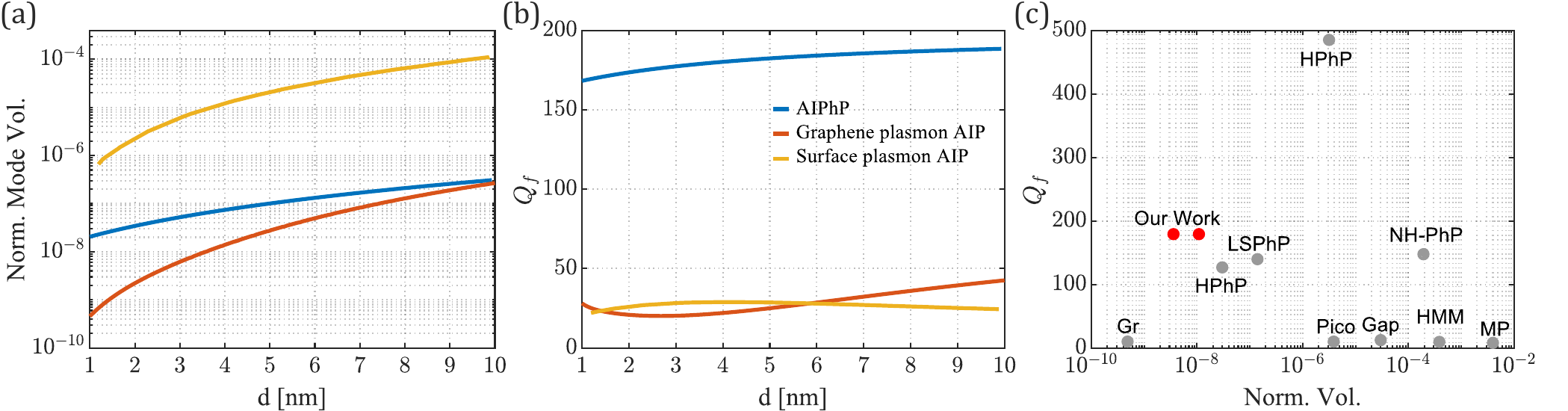}

               \caption{ (a) Normalized mode volume and (b) quality factor calculations as a function of the spacer thickness $d$, for AIPhPs (blue curve), graphene-plasmon AIPs (red curve), and surface-plasmon AIPs (orange curve) resonators. (c) Comparison to other type of resonators in terms of Q-factors vs. normalized cavity volume following \cite{HerzigSheinfux2024High-qualityNitride} : picocavities (Pico) \cite{Chikkaraddy2016Single-moleculeNanocavities}, nanogap plasmon polaritons (Gap) \cite{Kuttge2010UltrasmallResonators}, GPP (Gr) \cite{Epstein2020Far-fieldVolumes}, metallic particles (MP) \cite{Mock2002ShapeNanoparticles}, hyperbolic metamaterials (HMM) \cite{Yang2012ExperimentalLaws}, HPhP \cite{HerzigSheinfux2024High-qualityNitride},localized SPhP (LSPhP) \cite{Caldwell2013Low-lossResonators}, non hyperbolic PhP (NH-PhP) \cite{Wang2013OpticalPolaritons}. The two AIPhP points correspond to the calculated and physical mode volume.}
               \label{fig:5}

\end{figure}

Finally, we compute the mode volume and quality factor of the AIPhP resonators using the quasi-normal mode theory (see methods), and compare the obtained results to those surface-plasmon AIPs in the visible spectrum, and graphene-plasmons AIPs in the MIR spectrum. Figure \ref{fig:5}a presents the calculated normalized mode volume, $V_{polariton}/\lambda_0^3$, where $\lambda_0$ is the free-space wavelength, for each resonator with varying spacer thicknesses $d$ \cite{Moreau2012Controlled-reflectanceNanoantennas, Epstein2020Far-fieldVolumes}. It can be seen that our AIPhP resonator achieves a normalized mode volume approximately two orders of magnitude smaller than that of the surface-plasmon AIPs, reaching a value of $~10^{-8}$ for wavelengths that are over an order of magnitude larger. We note that while the graphene-plasmon AIPs exhibit smaller normalized mode volumes compared to the AIPhPs, their resonance reside shorter wavelengths.\par

Respectively, in Fig. \ref{fig:5}b we present the calculated Q-factor of the resonators as a function of spacer thickness. Notably, our AIPhP resonators achieve Q-factors that are an over an order-of-magnitude larger than the equivalent resonators, demonstrating its capacity to store a significantly larger amount of energy due to the reduced losses provided by SPhPs \cite{Lalanne2018LightResonances}. For completeness, we also compare our results with other types of resonators, however, since mode volume calculations of these are sparsely found in the literature, we follow the approach and terminology of Sheinfux et al.\cite{HerzigSheinfux2024High-qualityNitride}, taking the approximated mode volume as the physical volume of the resonator. The comparison is presented in Fig. \ref{fig:5}c, showing that our AIPhP resonators can achieve extremely small mode volumes with relatively large Q-factors.\par


\section{Conclusions}
In this study, we have measured and observed the far-field excitation of AIPhP resonators in SiC. We have demonstrated that each resonator exhibits remarkably small normalized mode volumes, with low losses, and with a spectrally controllable and tunable response. This was achieved by depositing silver NCs on top of a SiC surface, with each NC functioning as a AIPhPs resonator. This versatile platform may open the path for controlling and manipulating strong light-matter interactions at the nanoscale in the long-wavelength range.\par

\section{Methods}
\textit{Sample fabrication} - The 4H semi-insulating $500 um$ thick SiC wafers were obtained from MSE supplies and subsequently coated with single layers of poly-allamine hydrochloride (Sigma-Aldrich) followed by the deposition of silver NCs (NanoComposix), as described in \cite{Hoang2016ColloidalNanophotonics}. \par

\textit{Optical measurements} - The far-field reflectivity spectra of the samples were measured using a commercial FTIR (Bruker, Vertex V70) coupled to a microscope (Bruker, Hyperion) equipped with a 15X objective with a spectral resolution of $1cm^{-1}$. We conducted 400 scans for each FTIR measurement and smoothed each measurement using a Savitzky-Golay filter. The reflectance spectra reported were all in reference to a gold mirror. \par

\textit{Simulations} - Simulations of a single SiC-NC resonator were conducted to investigate non-periodic isolated objects utilizing the Rigorous Coupled-Wave Analysis (RCWA) method. This method, originally developed for periodic structures, was adapted for our specific application\cite{Lalanne2005PerfectlyFormalization}. The computation of poles in omega and the normalization of quasi-normal Modes (QNMs) are based on perfectly matched layers as nonlinear coordinate transforms\cite{Sauvan2013EfficientNanostructure}. Definitions of QNMs and modal volume were derived from relevant literature\cite{Zarouf2022NormalizationInvited}.\par

\section{Supporting Information}
Additional SEM images and simulation description. 

\section{Acknowledgments}
I.E. acknowledges the support of the Ministry of Science and Technology (MOST), via grant No. 0005757.

\bibliography{references}
\end{document}